\documentclass[aps,prb,twocolumn,showpacs]{revtex4}
\usepackage{graphicx}

\begin{document}

\title{Hydrogen-assisted distortion of gold nanowires}

\author{A.~Halbritter, Sz.~Csonka,  and G.~Mih\'aly}
\affiliation{Electron Transport Research Group of the Hungarian
Academy of Science and Department of Physics, Budapest University
of Technology and Economics, 1111 Budapest, Hungary}

\author{O.I.~Shklyarevskii, S.~Speller, and H.~van~Kempen}
\affiliation{NSRIM, University of Nijmegen, Toernooiveld 1, 6525
ED Nijmegen, the Netherlands}

\date{\today}

\begin{abstract}
In this paper the influence of adsorbed hydrogen on the behavior
of gold nanojunctions is investigated. It is found, that the
hydrogen environment has a strong effect on the conductance of
atomic-sized gold junctions, which is markedly reflected by the
growth of an additional peak in the conductance histogram at $0.5$
quantum unit, $2e^2/h$. The statistical analysis of the
conductance traces implies that the new peak arises due to the
hydrogen-assisted distortion of atomic gold chains.
\end{abstract}


\maketitle

\section{Introduction}

The conductance histograms of different metals have been studied
under various experimental conditions.\cite{Agrait2003} It was
found, that the peaks in the histogram arise due to the repeated
establishment of some stable atomic configurations during the
breaking process. Usually the general shape of the histogram is
characteristic of the material, and it is not very sensitive to
the experimental parameters, like the temperature or the magnitude
of the bias voltage.

On the other hand, the interaction of adsorbates with the junction
can influence the evolution of the nanocontact during the rupture
and can cause drastic changes in the shape of the histogram. It
was shown that the adsorption of some organic molecules on copper
nanojunctions results in the appearance of new peaks around $0.5$
and $1.5$ quantum conductance unit, G$_0=2e^2/h$.\cite{Li2000}
Similar phenomenon was observed using gold nanowires immersed in
an electrochemical cell.\cite{Shu2000} By changing the potential
on the nanoconstriction from positive to negative with respect to
the reference electrode, the authors observed the appearance and
growth of two new peaks, again at $0.5$ and $1.5$\,G$_0$.

These studies have shown that the conductance histograms can serve
as sensitive detectors of adsorbates. The new peaks in the
histogram are assumably caused by the appearance of new, stable
atomic arrangements in the modified environment. On the other
hand, the conductance histograms alone cannot tell the microscopic
background of the phenomenon. The above systems are rather
complicated for theoretical calculations, so the origin of the
observations has remained an open question.

In this paper we show that similar phenomenon can arise due to the
adsorption of hydrogen molecules on gold nanowires. Hopefully, the
simplicity of this system will open the opportunity for
theoretical studies. We also show, that the detailed statistical
investigation of the conductance traces supplies useful
information about the nature of the new atomic configurations.

Quite recently, the influence of hydrogen molecules on platinum
junctions was also studied.\cite{Smit2002} In this system the
characteristic peak of platinum at $1.5$\,G$_0$ disappears, and a
new peak grows in the histogram at $1$\,G$_0$. The detailed
studies have shown, that the new peak is connected to the
conduction through a H$_2$ bridge between the Pt electrodes. The
comparison of these results with our observations is
presented.

\section{Basic observation: the appearance of a new peak in the conductance
histogram}

\begin{figure}[t!]
\centering
\includegraphics[width=\columnwidth]{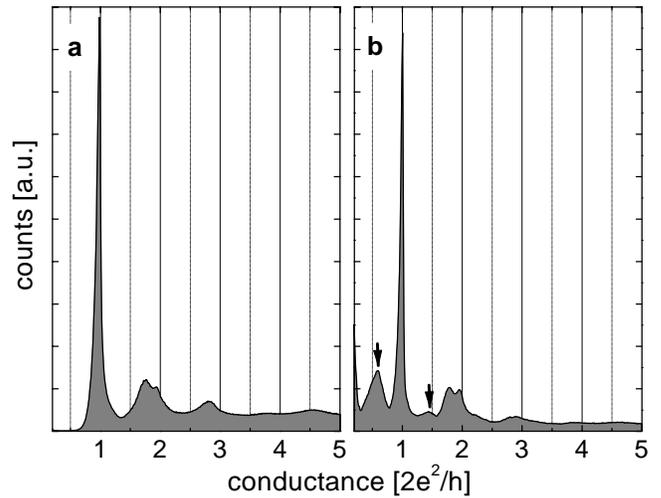}
\caption{\it Conductance histograms of gold measured in an
ultra-high vacuum environment (a) and in the presence of adsorbed
hydrogen (b). Both histograms were recorded at a bias voltage of
$129$\,mV and a temperature of $20$\,K. The histograms were built
from $10000$ conductance traces.} \label{AuH2hist.fig}
\end{figure}

We have performed our measurements on high-purity gold samples
with the mechanically controllable break junction (MCBJ)
technique. The experiments were done under cryogenic circumstances
in the temperature range of $4.2-50$\,K. A typical conductance
histogram of gold measured in ultra high vacuum at $T=20$\,K is
presented in Fig.~\ref{AuH2hist.fig}(a). It agrees with earlier
reported  data obtained under similar conditions.\cite{Yanson2001}
The most pronounced feature of this histogram is a sharp peak at
$1$\,G$_0$.  The introduction of hydrogen into the vacuum
pot\cite{note1} resulted in the appearance of an additional peak
positioned close to $0.5$\,G$_0$, as shown in
Fig.~\ref{AuH2hist.fig}(b). In some cases a small peak at $\sim
1.5$\,G$_0$ was also visible, but due to its eventual occurrence
we focus our attention on the peak at $\sim 0.5$\,G$_0$.

This basic observation was studied in detail by changing three
experimental parameters: the temperature, the bias voltage and the
amount of H$_2$ near the contact. Even though the quantity of gas
admitted into the vacuum pot was accurately determined, no
reliable estimation for the H$_2$ coverage of the junction can be
given due to the presence of different materials and unavoidable
temperature gradients in the sample holder. Fortunately, the
results of the measurements were not sensitive to the precise
amount. In contrast, the other two parameters played a rather
crucial role and the new peak in the histogram was only observed
in a restricted range of the temperature and the bias voltage.

\begin{figure}[b!]
\centering
\includegraphics[width=\columnwidth]{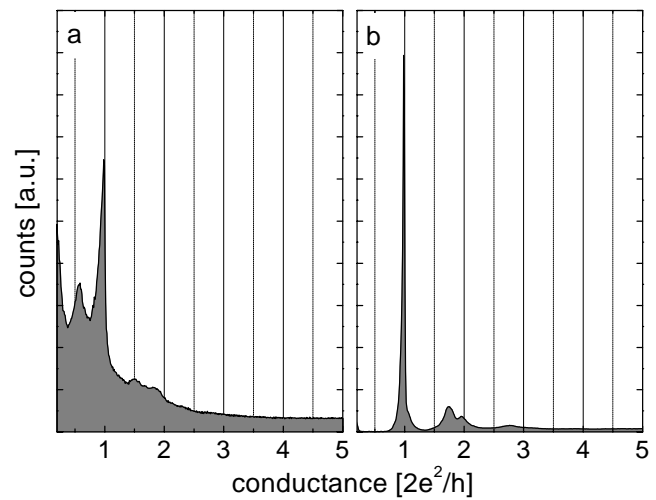}
\caption{\it Bias dependence of the conductance histograms of gold
measured at $T=20$\,K in hydrogen environment. The bias voltages
are $65$\,mV (a) and $240$\,mV (b). The histograms were built from
$10000$ conductance traces.} \label{AuH2hist2.fig}
\end{figure}

The study of the temperature dependence showed that the fractional
peak is only present in the range of approximately $10-30$\,K. The
bias dependence of the conductance histograms measured at a fixed
temperature ($20$\,K) is shown in Fig.~\ref{AuH2hist2.fig}. At
bias voltages $\lesssim 100$\,mV the peaks of the histogram are
superimposed on a large featureless background
[Fig.~\ref{AuH2hist2.fig}(a)]. At higher bias, both the background
and the relative amplitude of the peak at $0.5$\,G$_0$ compared to
the one at $1$\,G$_0$ decrease [Fig.~\ref{AuH2hist.fig}(b)]; and
finally above $\sim 200$\,mV the peak at $0.5$\,G$_0$ completely
disappears [Fig.~\ref{AuH2hist2.fig}(b)]. The absence of the new
peak at elevated temperature or bias voltage is presumably caused
by the desorption of hydrogen from the surface of the junction.
These two effects can have the same origin: at elevated bias the
voltage-induced heating of the junction\cite{Halbritter2002}
causes the desorption. At low temperatures the vapor pressure of
hydrogen is very small, so the amount of hydrogen near the contact
is strongly reduced.

The same measurements were performed on silver and copper
contacts. In high vacuum the histograms of these noble metals are
resembling the one of gold.\cite{Yanson2001} The inclusion of
hydrogen resulted in a similar behavior with respect to the
appearance of a featureless background at low bias voltage. On the
other hand, in Ag and Cu no indication for the appearance of a new
peak was observed.

Summarizing the experimental results, the addition of H$_2$ to the
contact area has two basic observable consequences: (1) the growth
of a large background signal at low bias voltage which is present
for all the materials investigated; (2) the appearance of a new
peak at $0.5$\,G$_0$ in the conductance histogram of gold.

The first phenomenon is presumably caused by a large amount of
H$_2$ between the electrodes. Due to the adsorption of hydrogen,
the surface energy of the junction is reduced. From this reason, a
larger amount of atomic configurations might occur during the
break. It is also possible that an increased tunnel current can
flow through the hydrogen layers.

The second phenomenon -- which is in the focus of the present
study -- indicates the repeated establishment of a special atomic
configuration during the last stages of the break. This atomic
configurations arises or becomes stable due to the surrounding
hydrogen molecules. The analysis of the conductance traces can
provide information about the nature of this new atomic
configuration, as discussed in the following.

\section{The nature of the conductance traces}

In pure gold contacts the conductance plateau at $1$\,G$_0$ is a
robust feature. The conductance traces contain this plateau almost
without exception. Furthermore, this is the last plateau, after
which the contact breaks. This plateau is very flat, very long,
and usually it is precisely positioned at $1$\,G$_0$. These
features are reflected by the exceptionally sharp peak in the
conductance histogram [Fig.~\ref{AuH2hist.fig}(a)]. There is some
chance for having smaller conductance values down to $0.6$\,G$_0$,
but below that there are absolutely no counts in the histogram.

\begin{figure}[b!]
\centering
\includegraphics[width=\columnwidth]{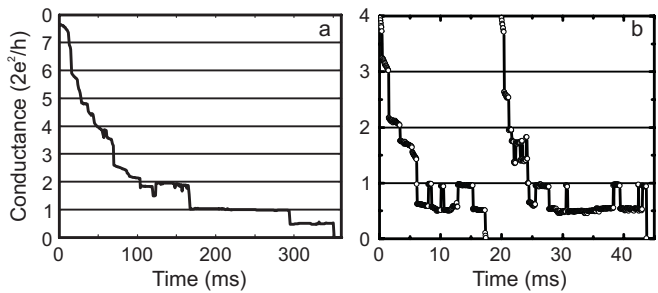}
\caption{\it Conductance traces of hydrogen-covered gold
junctions. Panel (a) shows a trace with a long plateau at
$1$\,G$_0$ followed by a shorter plateau at $0.5$\,G$_0$. The
traces in panel (b) show telegraph fluctuation between $1$\,G$_0$
and $0.5$\,G$_0$.} \label{AuH2traces.fig}
\end{figure}

The appearance of a new peak around $0.5$\,G$_0$ raises the
question, how these strong features of the conductance traces of
pure gold are modified in the presence of hydrogen. Naturally, the
traces must contain plateaus both at $0.5$\,G$_0$ and at
$1$\,G$_0$. However, the conductance histogram alone cannot tell
how these plateaus are related to each other. Figure
\ref{AuH2traces.fig}(a) shows a conductance trace extracted from
the data set. It has plateaus at both conductance values: first
the conductance stays at $1$\,G$_0$, then it jumps to
$0.5$\,G$_0$, and finally the contact breaks. This kind of
behavior was frequent, but not exclusive. Another two examples are
presented in Fig.~\ref{AuH2traces.fig}(b). These traces show a
telegraph fluctuation between $0.5$\,G$_0$ and $1$\,G$_0$. This
telegraph fluctuation implies that the contact can choose from two
metastable atomic configurations. The one with unit conductance is
assumably the customary atomic arrangement of pure gold. The other
one with $G=0.5$\,G$_0$ is another -- yet unknown -- configuration
which is only observable if hydrogen molecules are also present.
It should be noted that plateaus at $0.5$\,G$_0$ are not seen in
each trace, several conductance curves look like those in pure
gold contacts. On the other hand, traces with plateau at
$0.5$\,G$_0$ and no plateau at $1$\,G$_0$ were not typical.

The plateaus are not precisely placed at $1$\,G$_0$ and
$0.5$\,G$_0$, which is indeed expected from the finite width of
the peaks in the histogram. The peak at $0.5$\,G$_0$ grows above
the interval $\sim 0.3-0.75$\,G$_0$, while the peak at $1$\,G$_0$
grows above $\sim 0.75-1.1$\,G$_0$. So all the plateaus in the
first interval are regarded as plateaus near $0.5$\,G$_0$, while
the plateaus in the second interval are taken as plateaus near
$1$\,G$_0$.

Further on the lengths of these two types of plateaus are analyzed
with statistical methods.

\section{Exploring correlations between the peaks in the histogram}

\begin{figure}[t!]
\centering
\includegraphics[width=0.8\columnwidth]{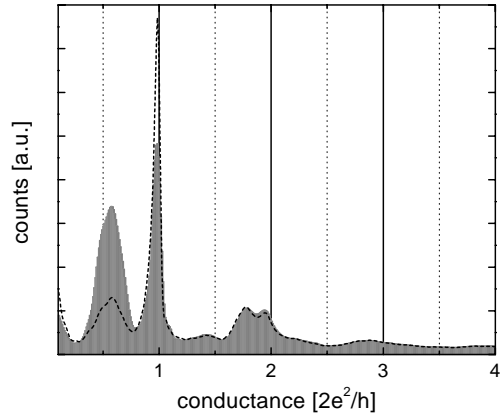}
\caption{\it The area graph shows the histogram for selected
$30$\,\% of the traces that have the longest plateau at $\sim
0.5$\,G$_0$. The dashed line shows the histogram for the original,
unselected data set. Both histograms are normalized to the number
of traces included.} \label{AuH2corrhist.fig}
\end{figure}

As a first step, we investigate those traces for that the plateau
at $0.5$\,G$_0$ is long. The length of this plateau was determined
for each trace by counting the number of data points in the
conductance interval $G/G_0 \in [0.3, 0.75]$. The data set of
Fig.~\ref{AuH2hist.fig}(b) contained $10000$ conductance traces.
From that those $3000$ traces were selected, which have the
longest plateau at $0.5$\,G$_0$. The histogram for these selected
curves is shown in Fig.~\ref{AuH2corrhist.fig} by the area graph.
This histogram is compared with the histogram for the whole data
set, as shown by the dashed line. In order to have a good
comparison, both histograms were normalized to the number of
traces included.

For the selected curves the peak at $0.5$\,G$_0$ is significantly
larger, which is a natural consequence of the selection. The
interesting result of this comparison is the shrinking of the peak
at $1$\,G$_0$ by more than $30$\,\% due to the selection. The rest
of the structures at higher conductance values are exactly the
same on the two histograms.

This analysis shows that the traces with long plateaus at
$0.5$\,G$_0$ have a smaller peak at $1$\,G$_0$ in the histogram;
or in other words, if the plateau at $0.5$\,G$_0$ is longer than
in average then the plateau at $1$\,G$_0$ is shorter than in
average. This kind of anticorrelation implies that the summed
length of the both plateaus is important, and the switching
between them is probably a random process. This remark will have
special importance in the next part, where the histograms for the
plateau lengths are investigated.

\section{Plateau length histograms}

The measurements in a pure environment have shown, that a single
atom gold contact has a single conductance channel with perfect
transmission.\cite{Scheer1998,Ludoph1999} Furthermore it was found
that gold can form atomic chains with single atoms in a
row,\cite{Yanson1998} and this chain also has a perfect
transmission.\cite{Rubio2003} The existence of these atomic chains
was deduced from histograms plotted for the lengths of the last
plateaus before break. These histograms revealed equidistant
peaks, corresponding to the break of chains with different number
of atoms. Therefore the question must be addressed, to what extent
this chain formation is influenced by the presence of hydrogen.

\begin{figure}[t!]
\centering
\includegraphics[width=0.6\columnwidth]{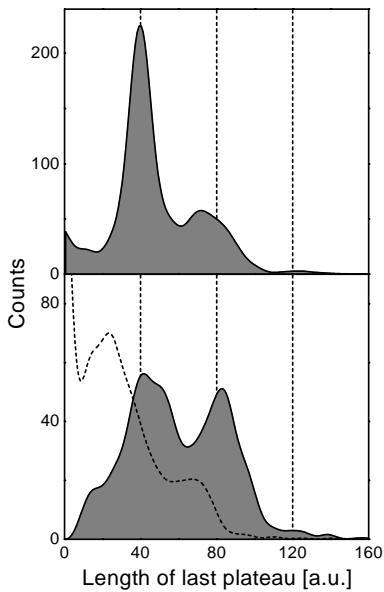}
\caption{\it Plateau length histograms for gold junctions in
hydrogen environment. Panel (a) shows the plateau length histogram
for the traces that do not have any plateau in the interval $G/G_0
\in [0.3, 0.75]$. The plateau length is determined by the number
of data points in the interval $G/G_0 \in [0.75, 1.1]$. Panel (b)
contains plateau length histograms for those traces that do have a
plateau in the region $G/G_0 \in [0.3, 0.75]$. The dashed line
shows the histogram for the lengths of the plateaus near
$1$\,G$_0$. This is measured by the number of points in the
interval $G/G_0 \in [0.75, 1.1]$. The area graph shows the
histogram for the joint lengths of the plateaus around $1$ and
around $0.5$\,G$_0$. This length is given by the number of points
in the interval $G/G_0 \in [0.3, 1.1]$} \label{AuH2plh.fig}
\end{figure}

As it was mentioned, even in hydrogen surrounding, a lot of traces
just exhibit plateaus at $1$\,G$_0$ without any additional plateau
around $0.5$\,G$_0$. These traces can be regarded as ``pure gold
behavior'', and the plateau length histogram for these traces can
serve as a reference [Fig.~\ref{AuH2plh.fig}(a)]. This histogram
shows two well-defined peaks. This means that the process of chain
formation is present, but no really long chains are pulled. It
should be noted, that in metals that cannot form chains the
plateau length histogram shows only a single peak. The distance
between the two peaks in Fig.~\ref{AuH2plh.fig}(a) defines the
Au-Au distance in the chain.

We can also plot the plateau length histogram for the traces that
have a plateau around $0.5$\,G$_0$ as well. The anticorrelation
demonstrated in the previous part implies that the summed length
of the plateaus at $1$\,G$_0$ and at $0.5$\,G$_0$ is important.
Therefore, as a first step the plateau length is measured for each
trace by the number of data points in the conductance interval
$G/G_0 \in [0.3, 1.1]$. The plateau length histogram built in this
way is presented by the area graph in Fig.~\ref{AuH2plh.fig}(b).
The position of the peaks coincides with the reference histogram.
The plateau length histogram for the same traces can be plotted by
measuring just the lengths of the plateaus at $1$\,G$_0$. (In this
case the plateau length is the number of data points in the
conductance interval $G/G_0 \in [0.75, 1.1]$) This analysis
results in the histogram shown by the dashed line. In this
histogram the peaks are smeared and considerably shifted compared
to the reference histogram.

A plateau at $0.5$\,G$_0$ in a conductance trace is a clear sign,
that the break of the nanojunction is influenced the hydrogen
molecules. The two peaks in the plateau length histogram in
Fig.~\ref{AuH2plh.fig}(b) indicate that even in this case the
chain formation of gold is preserved. Furthermore, a nice plateau
length histogram is only obtained if the plateaus at $0.5$\,G$_0$
are also included in the plateau length. This indicates that the
atomic configuration with $G=0.5$\,G$_0$ is a part of the chain
formation.

\section{Discussion of the observations}

Our measurements have shown, that the adsorption of hydrogen on
gold junctions results in the appearance of a new peak at
$0.5$\,G$_0$ in the conductance histogram. The same phenomenon was
not observed in silver and copper junctions, which have conduction
properties similar to gold. An important difference between the
nobel metals is the striking property that gold can form atomic
chains while Ag and Cu cannot.\cite{Smit2001} It has already
implied that the new peak at $0.5$\,G$_0$ is somehow connected to
the chain formation of gold. This assumption was further supported
by the investigation of plateau length histograms. This analysis
has shown that the configuration with $G=0.5$\,G$_0$ is indeed a
part of the chain formation.

Local density functional simulations have shown that a small
increase in the interatomic distance in a Au chain causes a
considerable reduction of the conductance.\cite{Hakkinen1999} This
implies, that the configuration with perfect transmission is the
usual gold chain, while the conductance value of $G=0.5$\,G$_0$
corresponds to a distorted chain with slightly increased Au-Au
distance.

This distortion can be an intrinsic property of gold, which is
just aided by the hydrogen environment. It was shown that
stretched atomic chains have a tendency for spontaneous
dimerization.
\cite{Torres1999,Okamoto1999,Hakkinen1999,Hakkinen2000,Maria2000}
Calculations of the conductance through a short chain of gold
atoms with a dimer yield values from $0.58$\,G$_0$
\cite{Hakkinen1999} to $0.4$\,G$_0$.\cite{Okamoto1999} Such dimers
have never been observed in high vacuum, but the adhesive hydrogen
environment might stabilize this configuration. A crucial problem
with this interpretation is the following. In a process called
``dimerization'' at least three atoms should be included in the
chain. In our measurements the occurrence of such long chains was
not frequent (see the plateau length histograms in
Fig.~\ref{AuH2plh.fig}), so a plausible explanation should also
account for the distortion of a chain with two atoms.

As another possibility, the binding of the hydrogen molecule on
the chain itself might cause the distortion. The precise nature of
such process cannot be predicted without microscopic model
calculations. The possibility of the chemical reaction of H$_2$
with the gold chain also cannot be excluded.

The measurements on hydrogen covered platinum junctions have
demonstrated that the new peak in the histogram is connected to
the conductance through a hydrogen molecule.\cite{Smit2002} In
gold the plateau length histograms strongly indicate that the
conductance values near $0.5$\,G$_0$ are related to chains of gold
atoms; therefore, the same explanation is not supported. A similar
analysis of the plateau length histograms can also be performed on
platinum, which has the tendency of chain formation as well. Our
measurements have shown that in Pt the chain formation is
completely destroyed in the presence of hydrogen.

In conclusion, we have shown that a new stable atomic
configuration appears in gold nanojunctions due to the adsorption
of hydrogen. The analysis of the results implies that this
configuration is related to a distorted chain of gold atoms.
Presently, the experimental observations cannot provide more
information about the nature of this configuration. Hopefully, the
simplicity of the gold-hydrogen adsorbate system will stimulate
extensive theoretical investigations, which will lead to a better
understanding of the details of molecular adsorption on nanowires.

This work has been supported by the ``Stichting FOM'' and the
Hungarian research funds OTKA TS040878, T037451.


\begin{thebibliography}{17}
\bibitem[{not()}]{note1}
\bibinfo{note}{The admitted amount of H$_2$ was between $1\mu$mole$-1$mmole.}
\end{thebibliography}
\end{document}